\documentclass[12pt]{article}
\usepackage{enumerate}
\usepackage{amsfonts}
\usepackage{amsmath}
\usepackage{amssymb}
\usepackage[active]{srcltx}
\usepackage{amsthm}

\def\H{\mathcal{H}}

\newcommand{\Tr}{\mathrm{Tr}}

\newcounter{defin}  \newcounter{lemma}  \newcounter{theorem}
\newcounter{property} \newcounter{corol}  \newcounter{remark} \newcounter{example}

\newenvironment{property}{\par\refstepcounter{property}     \textbf{Proposition \theproperty.}\ }{\rm\par}

\begin{document}

\title{Quantum conditional entropy for infinite-dimensional systems\thanks{
Work partially supported by RFBR grant 09-01-00424.}}
\author{A. A. Kuznetsova \\
Moscow State University\\
kelit@list.ru}
\date{}
\maketitle

\begin{abstract}
In this paper a general definition of quantum conditional entropy
for infinite-dimensional systems is given based on recent work of
Holevo and Shirokov \cite{H-Sh} devoted to quantum mutual and
coherent informations in the infinite-dimensional case. The
properties of the conditional entropy such as monotonicity,
concavity and subadditivity are also generalized to the
infinite-dimensional case.
\end{abstract}

\section{The definition}

In this paper a general definition of quantum conditional entropy
for infinite-dimensional systems is given based on recent work of
Holevo and Shirokov \cite{H-Sh} devoted to quantum mutual and
coherent informations in the infinite-dimensional case. The
necessity of such a generalization is clear in particular from the
study of Bosonic Gaussian channels and was stressed also in
\cite{FER}. We refer to \cite{H-Sh} for some notations and
preliminaries. Let $A,\dots $ be quantum systems described by the
corresponding Hilbert spaces $\mathcal{H}_{A},\dots $, and let
$\Phi $ be a channel from $A$ to $B$ with the Stinespring isometry
$V:\mathcal{H} _{A}\longrightarrow \mathcal{H}_{B}\otimes
\mathcal{H}_{E}$. Let $\rho _{AR}$ be a purification of a state
$\rho _{A}$ with the reference system $R$ and let $\rho _{BRE}$ be
the pure state obtained by action of the operator $V\otimes I_{R}
$. The \emph{mutual information} of the quantum channel $\Phi $ at
the state $\rho _{A}$ is defined similarly to finite-dimensional
case (cf. \cite{AS}, \cite{BNS}) as
\begin{equation*}
I(\rho _{A},\Phi )=H(\rho _{BR}\Vert \rho _{B}\otimes \rho _{R}),
\end{equation*}
and \emph{coherent information} at the state $\rho _{A}$
\emph{with finite entropy} $H(\rho _{A})$ is defined as follows \cite{H-Sh}
\begin{equation*}
I_{c}(\rho _{A},\Phi )=I(\rho _{A},\Phi )-H(\rho _{A}).
\end{equation*}
It is then shown that the above-defined quantity satisfies the
inequalities
\begin{equation*}
-H(\rho _{A})\leq I_{c}(\rho _{A},\Phi )\leq H(\rho _{A})
\end{equation*}
and the identity
\begin{equation}
I_{c}(\rho _{A},\Phi )+I_{c}(\rho _{A},\widetilde{\Phi })=0,  \label{basic2}
\end{equation}
where $\widetilde{\Phi }$ is the complementary channel.

Let  $A,C$ be two (in general, infinite-dimensional) systems and $\rho _{AC}$
a state such that $H(\rho _{C})<\infty .$ We define the \emph{conditional
entropy} as
\begin{equation}
H(C|A)=H(\rho _{C})-H(\rho _{AC}\Vert \rho _{A}\otimes \rho _{C}),
\label{other}
\end{equation}
where $H(\cdot \Vert \cdot )$ is the relative entropy which takes its values
in $[0,+\infty ],$ so that $H(C|A)$ is well defined as a quantity with
values in $[-\infty ,+\infty ).$ If in addition $H(\rho _{A})<\infty $ ,
one recovers the standard formula $H(C|A)=H(AC)-H(A).$

Now let $B$ be  infinite-dimensional system so that  arbitrary state $\rho
_{BC}$ can be purified to a pure state $\rho _{ABC}.$ Consider the channel $
\Phi =\mathrm{Tr}_{A}$ from $AB$ to $B$ so that $\widetilde{\Phi }=\mathrm{Tr
}_{B}$, then
\begin{equation}
I_{c}(\rho _{AB},\mathrm{Tr}_{B})\equiv I(\rho _{AB},\mathrm{Tr}_{B})-H(\rho
_{AB})=H(\rho _{AC}\Vert \rho _{A}\otimes \rho _{C})-H(\rho _{C})=-H(C|A).
\label{CEA}
\end{equation}
Similarly
\begin{equation}
I_{c}(\rho _{AB},\mathrm{Tr}_{A})\equiv I(\rho _{AB},\mathrm{Tr}_{A})-H(\rho
_{AB})=H(\rho _{BC}\Vert \rho _{B}\otimes \rho _{C})-H(\rho _{C})=-H(C|B).
\label{CEB}
\end{equation}
From the results of \cite{H-Sh} concerning the coherent information, we then
obtain that under the condition $H(\rho _{C})<\infty $ both $H(C|A),H(C|B)$
are well defined and satisfy
\begin{equation}
|H(C|A)|\leq H(\rho _{C}),\quad |H(C|B)|\leq H(\rho _{C}),  \label{bound}
\end{equation}
\begin{equation}
H(C|A)+H(C|B)=0,  \label{CAB}
\end{equation}

\section{Properties of conditional entropy}
\begin{property}
\emph{Let $\rho_{AB}$ be such a state that $H(\rho_A) < \infty.$
The function $H(A|B)$ has the following properties:}

1.monotonicity: \emph{the inequality
\begin{equation} \label{monot}
H(A|BC) \leq H(A|B).
\end{equation}
holds for any systems $A,B,C.$}

2. concavity in $\rho_{AB}:$ \emph{if  $\rho_{AB} = \alpha
\rho^1_{AB} + (1-\alpha)\rho^2_{AB}, \alpha \in [0,1]$, then
\begin{equation} \label{concavity}
 H(A|B)  \geq \alpha H(A^1|B^1) +
(1-\alpha)H(A^2|B^2).
\end{equation}}

3. subadditivity: \emph{the inequality
\begin{equation} \label{subadd}
H(AB|CD) \leq H(A|C) + H(B|D).
\end{equation}
holds for any systems $A,B,C,D$ such that $H(\rho_A) < \infty,$
$H(\rho_B) < \infty.$}
\end{property}

\textit{Proof.} 1. To prove monotonicity we rewrite the inequality
(\ref{monot}) using the definition (\ref{other}), i.e.
$$
H(\rho_A) - H(\rho_{ABC}\| \rho_{BC} \otimes \rho_A ) \leq
H(\rho_A) - H(\rho_{AB}\| \rho_B \otimes \rho_{A}).
$$
which is equivalent to
$$
H(\rho_{AB}\| \rho_B \otimes \rho_{A})  \leq H(\rho_{ABC}\|
\rho_{BC} \otimes \rho_{A}),
$$
and the last inequality holds by monotonicity of the relative
entropy with respect to taking the partial trace.

2. Let $P^A_n, P^B_k$ be arbitrary increasing sequences of finite
rank projectors in the spaces $\H_A, \H_B,$ strongly converging to
the operators $I_A, I_B$ respectively. Consider the sequence of
states
 $$
 \rho_{AB}^{nk} =  \lambda_{nk}^{-1} {(P_n^A \otimes P_k^B \rho_{AB} P_n^A \otimes P_k^B)} , \lambda_{nk} = {\Tr (P_n^A \otimes P_k^B)\rho_{AB}},
 $$
with the partial states $\rho_A^{nk}$, $\rho_B^{nk}$.

Let us show first that
\begin{equation}\label{convergence}
\lim_{n,k \rightarrow \infty}H(A_{nk}|B_{nk}) = H(A|B).
\end{equation}

By the definition (\ref{other}) $H(A_{nk}|B_{nk})  =
H(\rho_A^{nk}) - H(\rho_{AB}^{nk}\|\rho_B^{nk} \otimes
\rho_A^{nk}).$ By using lower semi-continuity of the von Neumann
entropy \cite{W}, we obtain
$$
\lim_{n,k \rightarrow \infty} \inf H(\lambda_{nk} \rho_A^{nk})
\geq H(\rho_A).
$$
On the other hand, \cite[lemma 4]{L} implies
$$
 H(\lambda_{nk} \rho_A^{nk}) = H(P_n^A (\Tr_B I \otimes P_k^B \rho_{AB} I \otimes
 P_k^B )P_n^A ) \leq H(\Tr_B I \otimes P_k^B \rho_{AB} I \otimes
 P_k^B ).
 $$
 Further, by the dominated convergence theorem for entropy \cite{W} we have
 $$
\lim_{n,k \rightarrow \infty} H(\Tr_B I \otimes P_k^B \rho_{AB} I
\otimes
 P_k^B ) = H(\rho_A),$$
since $\Tr_B (I \otimes P_k^B \rho_{AB} I \otimes  P_k^B) \leq
\rho_A$ and $H(\rho_A) < \infty.$
Thus by the theorem about the limit of the intermediate sequence
we obtain
$$ \lim_{n,k \rightarrow \infty} H(\lambda_{nk} \rho_A^{nk})=
H(\rho_A), \mbox { hence,} \lim_{n,k \rightarrow \infty}
H(\rho_A^{nk}) = H(\rho_A).
$$

Then we prove that
 $\lim\limits_{n,k \rightarrow \infty}
H(\rho_{AB}^{nk}\|\rho_A^{nk} \otimes \rho_B^{nk}) =
H(\rho_{AB}\|\rho_A \otimes \rho_B).$
Consider the following values
$$
 H_{nk} = H(\rho_{AB}^{nk} \| \rho_B^{nk}  \otimes
 \rho_A^{nk}) =  H(\rho_A^{nk}) + H(\rho_B^{nk}) -
H(\rho_{AB}^{nk}),
$$
$$
 \widetilde{H}_{nk} =  H\left(\rho_{AB}^{nk} \| {\eta_k^{-1}} {P_k^B \rho_B P_k^B } \otimes
 {\mu_n^{-1}} {P_n^A \rho_A
P_n^A }\right) =
   -
H(\rho_{AB}^{nk})  -
$$
$$
 - \Tr \left(\rho_A^{nk} \right) \log \left( {\mu_n^{-1}}  {P_n^A \rho_A P_n^A } \right) -  \Tr \left(\rho_B^{nk} \right) \log \left(  {\eta_k^{-1}} {P_k^B \rho_B P_k^B} \right),
$$
$$
\mbox{ where }\mu_n = \Tr P_n^A \rho_A,~ \eta_k = \Tr P_k^B \rho_B.
$$
By using again \cite[lemma 4]{L} we have
\begin{eqnarray*}
 \lim_{n,k \rightarrow \infty} \widetilde{H}_{nk}&=&\lim_{n,k \rightarrow
\infty} H\left(P_n^A \otimes P_k^B \rho_{AB} P_n^A \otimes P_k^B
\|P_k^B \rho_B P_k^B  \otimes P_n^A \rho_A P_n^A\right) \\
 &=& H(\rho_{AB} \| \rho_B \otimes
 \rho_A).
  \end{eqnarray*}
We will prove that  $\lim_{n,k \rightarrow \infty} H_{nk} =
H(\rho_{AB} \| \rho_A \otimes
 \rho_B)$ by considering the difference  $ H_{nk}-
 \widetilde{H}_{nk}$.
After some calculation we obtain that the difference tends to zero:
$$
 \lim_{n,k \rightarrow \infty} \left(H_{nk}- \widetilde{H}_{nk} \right) = \lim_{n,k \rightarrow \infty} \left( H \left( {\rho_A^{nk} } \|   {\mu_n^{-1}} {P_n^A \rho_A P_n^A } \right)  +
 H \left(  \rho_B^{nk}
 \| {\eta_k^{-1}} {P_k^B \rho_B P_k^B} \right) \right) =
 0.
$$
The last double limit is equal to zero since it follows from
\cite[lemma 4]{L} that
$$ 0 \leq  H \left( \lambda_{nk} {\rho_A^{nk} } \|  {P_n^A \rho_A P_n^A} \right) = H \left( P_n^A
 \Tr_B  (I_A \otimes P_k^B \rho_{AB} I_A \otimes P_k^B)  P_n^A \|P_n^A \rho_A
P_n^A  \right) \leq
$$
$$
 \leq H\left(\Tr_B (I_A \otimes P_k^B \rho_{AB} I_A \otimes P_k^B)\|\rho_A\right),
$$
and \cite[lemma 7]{H-Sh} implies
$$
 \lim_{n, k \rightarrow \infty} H\left(\Tr_B (I_A \otimes P_k^B
\rho_{AB} I_A \otimes P_k^B)\|\rho_A\right) = H(\rho_A\|\rho_A) =
0.
$$
Thus, the theorem about the limit of the intermediate sequence
implies
$$
 \lim_{n, k \rightarrow \infty} H \left( \lambda_{nk} {\rho_A^{nk} } \|
 {P_n^A \rho_A P_n^A} \right) =  \lim_{n, k \rightarrow \infty} H \left({\rho_A^{nk}} \|  {\mu_n^{-1}}{P_n^A \rho_A P_n^A }
 \right) = 0.
 $$
 Similarly we obtain that the second summand of the difference $H_{nk} - \widetilde{H}_{nk}$ also tends to zero:
$$ \lim_{n,k \rightarrow \infty} H \left( \rho_B^{nk} \| {\eta_k^{-1}} {P_k^B \rho_B P_k^B}
 \right) =  0.$$

Finally, we have
$$
\lim_{n,k \rightarrow \infty}H(\rho_{AB}^{nk}\|\rho_B^{nk} \otimes
\rho_A^{nk}) = H(\rho_{AB}\|\rho_B \otimes \rho_A), \mbox{ hence }
\lim_{n,k \rightarrow \infty}H(A_{nk}| B_{nk}) = H(A|B).
$$
To prove the concavity of the function H(A|B) let us consider
$$
\rho_{AB} = \alpha \rho^1_{AB} + (1-\alpha)\rho^2_{AB}, \alpha \in
[0,1].$$
 Also consider again
 $$ \rho_{AB}^{nk} =
{\lambda_{nk}^{-1}} {P^A_n \otimes P^B_k \rho_{AB} P^A_n \otimes
P^B_k} = $$ $$ = \frac {\alpha (P^A_n \otimes P^B_k \rho^1_{AB}
P^A_n \otimes P^B_k) + (1-\alpha)(P^A_n \otimes P^B_k \rho^2_{AB}
P^A_n \otimes P^B_k)} {\alpha \Tr (P^A_n \otimes P^B_k
\rho^1_{AB}) + (1-\alpha)\Tr(P^A_n \otimes P^B_k \rho^2_{AB})} =
\frac {\theta_{nk}^1 \rho_{AB}^{1nk} + {\theta_{nk}^2
\rho_{AB}^{2nk}} }{\theta_{nk}^1 + \theta_{nk}^2 },$$
$$~\mbox{где } \theta^1_{nk} = \alpha \Tr P^A_n \otimes P^B_k
\rho_{AB}^1, ~\rho_{AB}^{1nk} = \frac { \alpha P^A_n \otimes P^B_k
\rho_{AB}^1 P^A_n \otimes P^B_k} {\theta^1_{nk}},$$
$$\theta^2_{nk} = (1-\alpha) \Tr P^A_n \otimes P^B_k \rho_{AB}^2,~
\rho_{AB}^{2nk} = \frac { (1-\alpha) P^A_n \otimes P^B_k
\rho_{AB}^2 P^A_n \otimes P^B_k} {\theta^2_{nk}}.$$

 Due to concavity of the condition information on the finite rank states
we can write that
$$ H(A_{nk}|B_{nk}) \geq \frac {\theta_{nk}^1}
{\theta_{nk}^1+ \theta_{nk}^2} H(A_{nk}^1|B_{nk}^1) + \frac
{\theta_{nk}^2} {\theta_{nk}^1+ \theta_{nk}^2}
H(A_{nk}^2|B_{nk}^2)  $$ Taking the limit and using
(\ref{convergence}) for both parts of inequality we obtain the
assertion of (\ref{concavity}). Thus, concavity is proved.

3. A direct verification shows that in the finite-dimensional case
\begin{equation} \label{subadd_1}
H(AB|CD) = H(A|CD) + H(B|CD) - (H(A|CD)-H(A|BCD)),
\end{equation}
which implies
\begin{equation} \label{subadd_2}
H(AB|CD) \leq H(A|CD) + H(B|CD),
\end{equation}
since the value in brackets in (\ref{subadd_1}) is nonnegative. We
will prove the inequality (\ref{subadd_2}) in infinite-dimensional
case, the subadditivity property (\ref{subadd}) can be derived
from (\ref{subadd_2}) by using monotonicity of the conditional
entropy.

Let $\rho_{ABCD} = \rho \in \H_A\otimes \H_B \otimes \H_C \otimes
\H_D$ be a state with $H(\rho_A) < \infty, H(\rho_B) < \infty.$
Consider  an arbitrary increasing sequence of finite rank
projectors $\{P^{CD}_l\} $ which strongly converges to the
operator $I_{CD}$.
 Also consider the sequence of states
$$
  \rho^{l} = {\tau_{l}^{-1}} {(I_A \otimes I_B \otimes P_l^{CD})\rho (I_A \otimes I_B \otimes P_l^{CD})} , \tau_{l} = {\Tr (I_A \otimes I_B \otimes
 P_l^{CD})\rho}.
 $$
The relation (\ref{subadd_1}) (which implies (\ref{subadd_2}))
holds for $\rho^l$ since all the summands in (\ref{subadd_1}) are
finite.

We will show that
$
H(A^{l}|CD^{l})\rightarrow H(A|CD).
$
Actually, using the definition (\ref{other}) we have
$$
H(A^{l}|CD^{l}) = H(\rho_A^l) - H(\rho_{ACD}^l\|\rho_{CD}^l
\otimes \rho_A^l).
$$
It follows from the dominated convergence theorem for entropy
\cite{W} that
 $$ \lim_{l \rightarrow
\infty} H(\rho_A^l) = \lim_{l \rightarrow \infty} H \left(
{\tau_{l}^{-1}}  { \Tr_{BCD} (I_A \otimes I_B \otimes P^{CD}_l)
\rho }  \right) = H(\rho_A).
$$
Further, consider the values
$$ H_l= H(\rho_{ACD}^l\|\rho_{CD}^l
\otimes \rho_A^l) = - H(\rho_{ACD}^l) + H(\rho_A^l) +
H(\rho_{CD}^l).
$$
and
$$ \widetilde{H}_l = H(\rho_{ACD}^l\|\rho_{CD}^l \otimes \rho_A) = H\left({\tau_{l}^{-1}} {I_A \otimes P^{CD}_l [\rho_{ACD}] I_A \otimes
P^{CD}_l} \| \rho_A \otimes {\tau_{l}^{-1}}  {P^{CD}_l [\rho_{CD}]
P^{CD}_l}
 \right) = $$ $$ =- H(\rho_{ACD}^l) + H(\rho_{CD}^l) + \Tr
\left( {\tau_{l}^{-1}}  {\Tr_{CD} I_A \otimes P^{CD}_l
[\rho_{ACD}]} \right) (-\log \rho_A).$$

Then \cite[lemma 4]{L} implies that
$$
\lim_{l \rightarrow \infty} \widetilde{H}_l = H(\rho_{ACD}\|\rho_A
\otimes \rho_{CD}).
$$
On the other hand, after the calculation
 and using  \cite[lemma 7]{H-Sh}
 we obtain
$$
\lim_{l \rightarrow \infty} (\widetilde{H}_l - H_l) = \lim_{l
\rightarrow \infty} H \left( {\tau_{l}^{-1}} {\Tr_{CD} I_A \otimes
P^{CD}_l [\rho_{ACD}]} \| \rho_A \right) = 0.
$$
This implies that
$$ \lim_{l \rightarrow \infty} H(\rho_{ACD}^l\|\rho_A^l \otimes
\rho_{CD}^l) = H(\rho_{ACD}\|\rho_A \otimes \rho_{CD}), \mbox {
hence,} \lim_{l \rightarrow \infty} H(A^l|CD^l) = H(A|CD).$$
In the similar way we obtain $H(B^{l}|CD^{l}) \rightarrow H(B|CD)$
and \\$H(AB^{l}|CD^{l}) \rightarrow H(AB|CD).$ Thus the statement
(\ref{subadd_2}) is proved  in infinite-dimensional case, hence,
the subadditivity property holds. $\quad \Box$

The following observation is due to M. E. Shirokov.

\begin{property}\label{continuity}
\emph{Let $\mathcal{A}$ be a subset of
$\mathfrak{S}(\mathcal{H}_{AC})$ such that the von Neumann entropy
is continuous on the set
$\mathcal{A}^{C}=\mathrm{Tr}_{A}\mathcal{A}\subset\mathfrak{S}(\mathcal{H}_{C})$.
Then the function $\rho_{AC}\mapsto H(C|A)$ is continuous on the
set $\mathcal{A}$. }
\end{property}

\begin{proof}
Let $\{\rho_{n}\}\subset\mathcal{A}$ be a sequence converging to a
state $\rho_{0}\in\mathcal{A}$. By the well known results of
purification theory there exists a corresponding sequence of
purifications
$\{\hat{\rho}_{n}\}\subset\mathfrak{S}(\mathcal{H}_{ABC})$
converging to a purification
$\hat{\rho}_{0}\in\mathfrak{S}(\mathcal{H}_{ABC})$ of the state
$\rho_{0}$. The sequence $\{\mathrm{Tr}_{C}\hat{\rho}_{n}\}$
converges to $\mathrm{Tr}_{C}\hat{\rho}_{0}$ and
$\lim_{n\rightarrow+\infty}
H(\mathrm{Tr}_{C}\hat{\rho}_{n})=H(\mathrm{Tr}_{C}\hat{\rho}_{0})$
by the condition (since
$H(\mathrm{Tr}_{C}\hat{\rho}_{n})=H(\mathrm{Tr}_{AB}\hat{\rho}_{n})$).
Proposition 4 in \cite{H-Sh} implies
$$
\lim_{n\rightarrow+\infty}I_c
(\mathrm{Tr}_{C}\hat{\rho}_{n},\Tr_B)=I_c
(\mathrm{Tr}_{C}\hat{\rho}_{0},\Tr_B).
$$
\end{proof}

The author is grateful to A.S. Holevo and M.E. Shirokov for suggesting the problem and for
discussions and the help in preparing the manuscript.


\begin{thebibliography}{9}
\bibitem{AS} \textit{Adami C., Cerf N.J.} Capacity of noisy quantum channel //
Phys.Rev. A. 1997. V.56 P.3470-3485; arXiv: quant-ph/9609024.

\bibitem{BNS} \textit{Barnum H., Nielsen M.A. Schumacher B.} Information transmission
through a noisy quantum channel // Phys. Rev. A. 1998. V.57
P.4153- 4175; arXiv:quant-ph/9702049.

\bibitem{H-Sh} \textit{Holevo A. S., Shirokov M. E.} Mutual and coherent
informations for infinite-dimensional quantum channels.
arXiv:1004.2495[quant-ph].

\bibitem{W} \textit{Wehrl A.} General properties of entropy// Reviews of
Modern Physics. 1978. V.50, N.2, P. 221-260.

\bibitem{L} \textit{Lindblad G.} Expectations and Entropy Inequalities for
Finite Quantum Systems// Commun. Math. Phys. 1974. V. 39. P. 111-119.

\bibitem{H-Sh-2} \textit{Holevo A. S., Shirokov M. E.} On approximation of
infinite-dimensional quantum channels.
Probl. Inform. Transmission. 2008. v. 44. n. 2. p. 3-22;
arXiv:0711.2245[quant-ph].

\bibitem{FER} \textit{Fabian Furrer, Johan Eberg, and Renato Renner}
Min- and Max-Entropy in Infinite Dimensions //arXiv:1004.1386[quant-ph].

\end{thebibliography}
\end{document}